\pdfoutput=1
\documentclass[11pt,twoside,a4paper,cmspaper,final,collab]{cms-tdr}
\def\svnVersion{881188b}\def\svnDate{2023/07/20}\def\cmsCernNoTag{CERN-EP-2023-014}\def\cmsCernDate{\today}\def\cmsMessage{Published in Physics Letters B as \href{http://dx.doi.org/10.1016/j.physletb.2023.138076}{\doi{10.1016/j.physletb.2023.138076}.}}
\begin{document}\cmsNoteHeader{TOP-21-005}

\newcommand{\tttt}{\ensuremath{\PQt{}\PAQt{}\PQt{}\PAQt}\xspace}
\newcommand{\pp}{\ensuremath{\Pp\Pp}\xspace}
\newcommand{\nj}{\ensuremath{N_\text{j}}\xspace}
\newcommand{\nb}{\ensuremath{N_\PQb}\xspace}
\newcommand{\nrt}{\ensuremath{N_\text{RT}}\xspace}
\newcommand{\nbt}{\ensuremath{N_\text{BT}}\xspace}
\newlength\cmsTabSkip\setlength{\cmsTabSkip}{1ex}
\ifthenelse{\boolean{cms@external}}{\providecommand{\cmsLeft}{upper\xspace}}{\providecommand{\cmsLeft}{left\xspace}}
\ifthenelse{\boolean{cms@external}}{\providecommand{\cmsRight}{lower\xspace}}{\providecommand{\cmsRight}{right\xspace}}

\cmsNoteHeader{TOP-21-005}
\title{Evidence for four-top quark production in proton-proton collisions at \texorpdfstring{$\sqrt{s} = 13\TeV$}{sqrt(s) = 13 TeV}}

\abstract{The production of four top quarks (\tttt) is studied with LHC proton-proton collision data samples collected by the CMS experiment at a center-of-mass energy of 13\TeV, and corresponding to integrated luminosities of up to 138\fbinv. Events that have no leptons (all-hadronic), one lepton, or two opposite-sign leptons (where lepton refers only to prompt electrons or prompt muons) are considered. This is the first \tttt measurement that includes the all-hadronic final state. The observed significance of the \tttt signal in these final states of 3.9 standard deviations (1.5 expected) provides evidence for \tttt production, with a measured cross section of $36^{+12}_{-11}\unit{fb}$. Combined with earlier CMS results in other final states, the signal significance is 4.0 standard deviations (3.2 expected). The combination returns an observed cross section of $17 \pm 4 \stat \pm 3 \syst\unit{fb}$, which is consistent with the standard model prediction.\par\vspace{2ex}\noindent\textit{We dedicate this publication to our friends and colleagues Meenakshi Narain and Stephen Wimpenny, who passed away unexpectedly while this paper was in preparation. This work would not have been possible without their guidance and contributions.}}

\hypersetup{
pdfauthor={CMS Collaboration},
pdftitle={Evidence for four-top quark production in proton-proton collisions  at sqrt{s} = 13 TeV},
pdfsubject={CMS},
pdfkeywords={CMS, top quark}}

\maketitle 

\section{\label{sec:intro}Introduction}

\begin{figure*}[htbp]
\centering
\includegraphics[width=.25\textwidth]{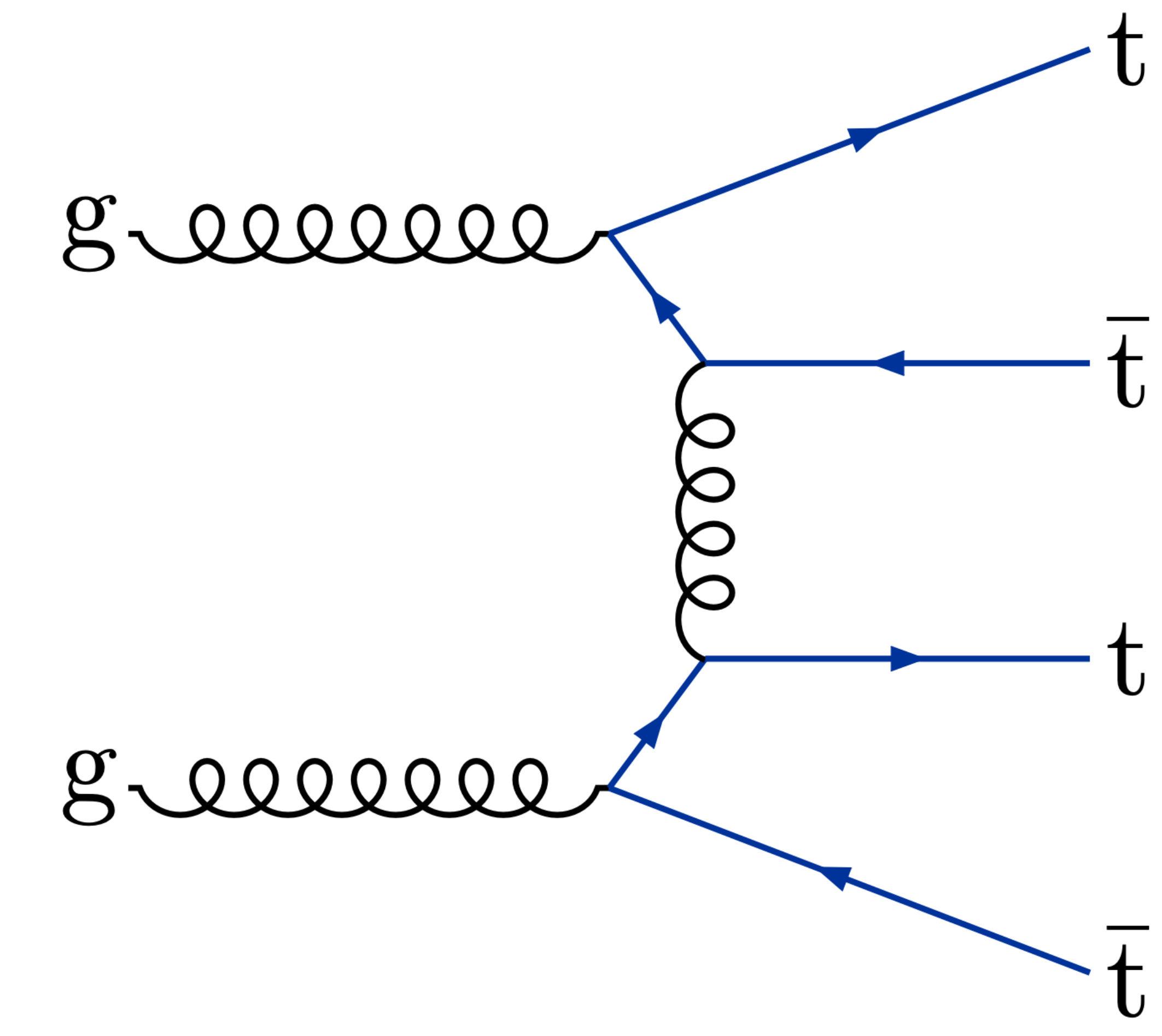}
\hspace{0.10\textwidth}
\includegraphics[width=.25\textwidth]{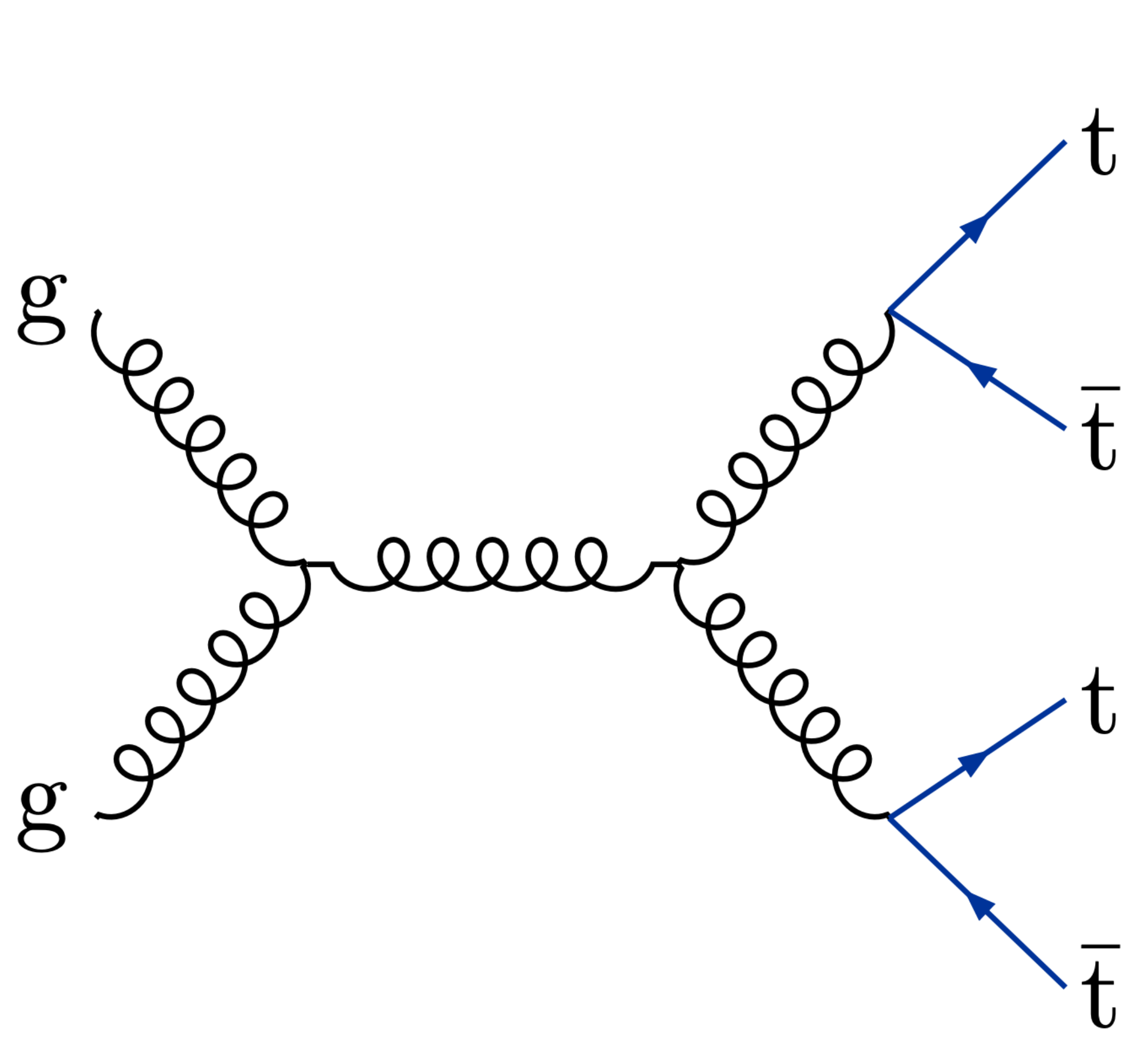}
\hspace{0.10\textwidth}
\includegraphics[width=.25\textwidth]{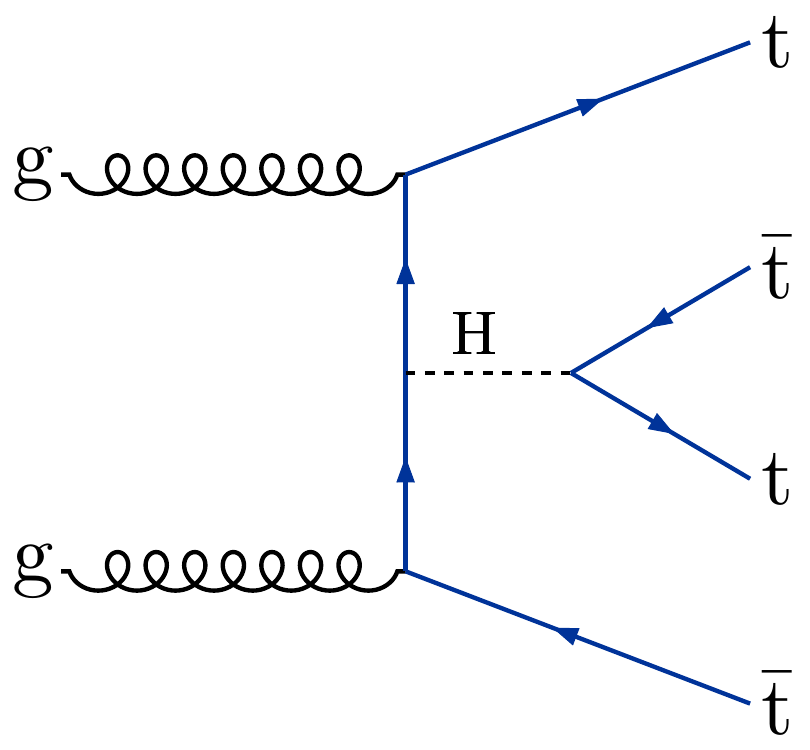}
\caption{Examples of Feynman diagrams for \tttt production at leading order in the SM.}
\label{fig:feyndiagrams}
\end{figure*}

The production of four top quarks (\tttt) is predicted to occur very rarely in the standard model (SM).
In proton-proton (\pp) collisions at $\sqrt{s}=13\TeV$, the production cross section has been calculated to be $12.0^{+2.2}_{-2.5}\unit{fb}$ at next-to-leading order (NLO) in quantum chromodynamics (QCD) and electroweak (EW) contributions~\cite{Alwall:2014hca,Jezo:2021smh,Frederix_2018,Bevilacqua:2012em,Frederix:2018nkq}.
Examples of the SM lowest order contributions to \tttt production in pp collisions are shown in Fig.~\ref{fig:feyndiagrams}.

Deviations from the predicted value occur in many proposed models of physics beyond the SM, such as supersymmetry~\cite{Nilles:1983ge,Martin_1997ns}, composite models~\cite{Cacciapaglia:2015eqa}, top quark compositeness~\cite{Kumar:2009vs}, two Higgs doublet models~\cite{Dicus:1994bm,Craig:2015jba,Craig:2016ygr}, and models with extra spatial dimensions~\cite{Cacciapaglia:2009pa,Ducu:2015fda}.
Measurements of \tttt production can also be used to constrain the top quark Yukawa coupling, $CP$-related parameters, and effective field theory operators~\cite{Degrande:2010kt}.

In the SM, the top quark dominantly decays to a bottom quark and a \PW boson.
Each \PW boson decays to either leptons or quarks, so the \tttt final state consists of four bottom quarks and up to four leptonic \PW boson decays. 
No analyses described or cited in this Letter attempt to explicitly identify $\tau$ lepton products and hence hereafter ``lepton'' will refer only to \Pe and \PGm, whether produced directly in $\PW\to \Pe/\PGm+\PGn$ decays or via $\PW\to\PGt + \nu$ with $\PGt\to \Pe/\PGm+2\PGn$.

The production of \tttt has been searched for by both ATLAS~\cite{Aaboud_tttt2016,Aaboud_tttt2016ss, ATLAS:2020hpj} and CMS~\cite{SS2016,tttt2016, Sirunyan:2019wxt,Sirunyan:2019nxl} at the CERN LHC.
The ATLAS Collaboration has reported evidence for \tttt production  in final states with either two same-sign leptons or at least three leptons  (referred to as ``SSDL\&ML'')~\cite{ATLAS:2020hpj},
with an observed significance of 4.3 standard deviations (2.4 expected) and a measured production cross section of $24^{+7}_{-6}\unit{fb}$ (assuming
SM branching ratios).
The significance observed in searches for \tttt production by the CMS Collaboration in similar SSDL\&ML final states is 2.6 standard deviations (2.7 expected), with a measured production cross section of $12.6^{+5.8}_{-5.2}\unit{fb}$, using data collected in 2016--2018 with an integrated luminosity of 138\fbinv~\cite{Sirunyan:2019wxt}.
The CMS Collaboration has also reported an observed significance of 1.4 standard deviations and a measured production cross section of $13^{+11}_{-9}\unit{fb}$ in final states with one lepton or two opposite-sign leptons, using data collected in 2016
with an integrated luminosity of 38\fbinv~\cite{Sirunyan:2019nxl}.

In this Letter, we present a search by the CMS Collaboration for the production of \tttt in final states with zero leptons (all-hadronic), one lepton (single-lepton), or two opposite-sign leptons (referred to as ``opposite-sign dileptons'' or ``OSDL'').
The search uses data samples of \pp collision data collected in 2016--2018 at $\sqrt{s}=13\TeV$, with integrated luminosities of up to 138\fbinv~\cite{CMS-LUM-17-003,CMS-PAS-LUM-17-004,CMS-PAS-LUM-18-002}.
To discriminate the \tttt signal events from the dominant background of \ttbar production we take advantage of the higher multiplicity of jets, particularly those produced by the hadronization of \PQb quarks.
Events are required to have at least three candidate \PQb jets in the single-lepton and all-hadronic final states, and at least two candidate \PQb jets in the OSDL final state.

The all-hadronic final state was not used in previous measurements because of the difficulty of modeling the QCD multijet background in the relevant kinematic regions.
Utilizing this channel for the first time is made possible by a novel background estimation strategy that uses a deep neural network (DNN) to estimate background distributions from data in control regions (CRs).
The analyses of channels with leptons in the final state improve upon previous results by taking advantage of the increased integrated luminosity, upgraded detectors, and improved algorithms for the identification of bottom and top quarks.
The results are combined with the previous SSDL\&ML CMS results~\cite{Sirunyan:2019wxt}, which are also based on the 2016--2018 data set but use non-overlapping data regions (including CRs) to ensure statistical independence. 
Tabulated results are provided in the HEPData record for this analysis~\cite{hepdata}.

\section{\label{sec:det}The CMS detector}

The central feature of the CMS apparatus is a superconducting solenoid providing a magnetic field of 3.8\thinspace T, and enclosing a silicon pixel and strip tracker, a lead tungstate crystal electromagnetic calorimeter, and a brass and scintillator hadron calorimeter, each composed of a barrel and two endcap sections. Forward calorimeters extend the pseudorapidity coverage provided by the barrel and endcap detectors. Muons are measured in gas-ionization detectors embedded in the steel flux-return yoke outside the solenoid.
A two-level trigger system reduces the rate of events retained for further processing to around 1\unit{kHz}.
The first-level trigger is composed of custom hardware processors, using information from the calorimeters and muon detectors~\cite{Sirunyan:2020zal}.
The software-based high-level trigger~\cite{Khachatryan:2016bia} uses the full event information.
A more detailed description of the CMS detector, together with a definition of the coordinate system used and the relevant kinematic variables, can be found in Ref.~\cite{Chatrchyan:2008zzk}.

\section{\label{sec:sim}Simulated event samples}

Signal and background processes are modeled using several Monte Carlo event generators.
Multiple simulated minimum bias interactions within the same or nearby bunch crossings (pileup) are superimposed on the hard scattering process with the multiplicity distribution matched to the  data.
Signal events are generated using the \MGvATNLO generator~\cite{Alwall:2014hca} versions 2.2.2 (2016) and 2.4.2 (2017--2018) in a tree-level approximation, with
emission of up to two additional partons in matrix element calculations.
Inclusive \ttbar production is generated at NLO precision using \POWHEG~v2~\cite{Nason:2004rx,Frixione:2007vw,Alioli:2010xd}. 
Smaller contributions from \ttbar production in association with one or two bosons (\PH, \PW, \PZ, {\PW}\PH, {\PZ}\PH, {\PW}\PW, {\PW}\PZ, {\PZ}\PZ), \PW and \PZ production, triple top quark production ({\ttbar}\PQt, {\ttbar}{\PQt}\PW), single top quark ({\PQt}\PW) production, and Drell--Yan processes constitute the remaining backgrounds, particularly where additional hadronic jets are produced by QCD radiation~\cite{Sirunyan:2019wxt}.
Single top and \ttbar + \PH processes are simulated using \POWHEG at NLO, while the other smaller contributions are simulated using \MGvATNLO versions 2.2.2 (2016) and 2.4.2 (2017--2018) at leading order (LO).

Fragmentation and parton showering are modeled by \PYTHIA~8~\cite{Sjostrand:2014zea} versions 8.212 (2016) and 8.230 (2017--2018). For samples generated with LO (NLO) precision, the MLM~\cite{Alwall:2007fs} (FxFx~\cite{Frederix:2012ps}) matrix element to parton shower matching scheme is used. 
For the 2016 simulation, the underlying event tunes CUETP8M1~\cite{CMS:2015wcf} and CUETP8M2T4~\cite{CMS-PAS-TOP-16-021} are applied to Drell--Yan and \ttbar production in association with two bosons, respectively.
The tune CP5~\cite{Sirunyan:2019dfx} is applied to the remaining processes in the 2016 simulation and all processes in the 2017--2018 simulation.
Simulations for the 2016 data-taking conditions are generated using the NNPDF3.0~\cite{Ball:2014uwa} parton distribution functions (PDFs) with either LO or NLO accuracy, while for 2017--2018 samples NNPDF3.1~\cite{Ball:2017nwa} at next-to-NLO (NNLO) accuracy is used.

Cross sections at NLO and NNLO are used to normalize the simulated background samples.
Events in the \ttbar sample with one or more additional \PQb jets from initial- or final-state radiation are reweighted to match corresponding cross sections measured by the CMS experiment: Ref.~\cite{Sirunyan:2020kga} is used for the SL analysis and Ref.~\cite{CMS:2020utv} for the OSDL analysis.
The detector response is simulated using \GEANTfour~\cite{Agostinelli:2002hh}.
In the single-lepton and OSDL final states, the contributions from all background processes are estimated using simulation.
In the all-hadronic final state, CRs in data are used to estimate the dominant multijet and \ttbar backgrounds.

\section{\label{sec:reco}Event reconstruction and data samples}

A particle-flow algorithm~\cite{CMS-PRF-14-001} is used to reconstruct and identify each particle in an event, with an optimized combination of information from various CMS subdetectors. 
The objects identified by the algorithm comprise candidate electrons, muons, photons, and charged or neutral hadrons.
Muon and electron candidates are restricted to the ranges $\abs{\eta} < 2.4$ and  $\abs{\eta}<2.5$, respectively~\cite{CMS:2020uim,CMS-DP-2020-021,CMS:2018rym},
and are required to be isolated from other objects~\cite{CMS:2020uim,CMS:2018rym,Rehermann:2010vq}.
Jets are reconstructed from particle flow objects using the anti-\kt algorithm~\cite{Cacciari:2008gp, Cacciari:2011ma} with distance parameters of 0.4 (AK4) and 0.8 (AK8).
Residual differences in the jet energy scale and resolution between data and simulation are corrected~\cite{Khachatryan:2016kdb}.
Charged particles identified as originating from pileup are discarded, and the measured energy is corrected to remove the estimated contribution from neutral pileup particles~\cite{Bertolini:2014bba,CMS:2020ebo}.
The \HT in an event is then defined as the scalar \pt sum of all AK4 jets with $\pt > 30\GeV$ and $\abs{\eta} < 2.5$.

The \textsc{DeepCSV}~\cite{DeepCSV} and \textsc{DeepJet} algorithms~\cite{Bols:2020bkb} are used to discriminate (\PQb tag) AK4 jets produced by the hadronization of bottom quarks from those produced by gluon and lighter quark hadronization.
A misidentification probability of 1\% and efficiencies of approximately 68\% for \textsc{DeepCSV} and 75\% for \textsc{DeepJet}
are measured in simulated \ttbar events~\cite{CMS-DP-2018-058}.

Hadronic top quark decays with a small or moderate Lorentz boost will typically produce three separate AK4 jets. These decays are referred to as ``resolved'' and are identified using  multivariate top quark tagging (\PQt~tagging) algorithms.
For the analysis of the single-lepton final state we use the DNN-based \textsc{DeepResolved} \PQt~tagger~\cite{Sirunyan:2019glc, CMS:2021beq} while for the all-hadronic final state we use a custom boosted decision tree (BDT) \PQt tagger~\cite{Hastie-Tibshirani-Friedman,Breiman-Friedman-Olshen-Stone} based on the one described in Ref.~\cite{Sirunyan_2017}, updated to use \textsc{DeepJet} \PQb tagging inputs.

In contrast, hadronically decaying top quarks with a large boost can result in a single merged jet.
Such decays are identified in the all-hadronic channel by applying the CMS \textsc{DeepAK8} algorithm~\cite{CMS-PAS-JME-18-002} to AK8 jets with $\pt>400\GeV$.
In order to avoid double counting of objects, we require merged top quark jets to be separated from resolved top quark candidates by $\DR \equiv \sqrt{\smash[b]{(\Delta\eta)^2 + (\Delta\phi)^2} } > 0.8$. 

The OSDL events are collected using electron-muon, dimuon, and dielectron triggers.
The $\Pe\Pgm$ selection uses a combination of triggers that require either an electron with $\pt>23\GeV$ and a muon with $\pt>12\GeV$, or vice-versa.
For the dimuon channel, a trigger with \pt thresholds of 17 and 8\GeV for the two highest \pt muons is used. Similarly the dielectron channel uses a trigger with \pt thresholds of 23 and 12\GeV for the two highest \pt electrons.
A prioritized trigger strategy of $\Pe\Pgm$, $\Pgm\Pgm$, and $\Pe\Pe$ is used to ensure that events satisfying multiple triggers exclusively enter the appropriate OSDL final state.
For the single-lepton channel, two different triggers are used.
The first requires events to contain an isolated electron (muon) with $\pt > 35\,(50)\GeV$. The second requires a very loosely isolated electron or muon with $\pt > 15\GeV$ in addition to the event having $\HT>450\GeV$.
All-hadronic events are selected with a variety of triggers that require at least six AK4 jets and \HT greater than thresholds in the range 380--450\GeV.
A minimum of either one or two \PQb-tagged jets is required, depending on the trigger.
Scale factors are applied to simulated samples in all final states to correct for the differences in trigger efficiencies between data and simulation.

\section{\label{sec:OSdilepton}Opposite sign dilepton final state}

The OSDL channel contains data corresponding to an integrated luminosity of 101\fbinv collected in 2017--2018, with previously published results on 2016 data included in the final fit to all channels.
Offline, events in the OSDL channel are required to have exactly two opposite-sign leptons, one with $\pt > 25\GeV$ and the other with $\pt > 15\GeV$, at least 4 jets with $\pt > 30\GeV$, and $\HT>500\GeV$.
Events with $\Pe\Pe$ and $\Pgm\Pgm$ invariant masses below 20\GeV or within 15\GeV of the \PZ boson mass are excluded.
The distribution of \HT is chosen as the input to the final fit, as this variable is sensitive to the presence of the extra two hadronic top quark decays in the signal as compared to the \ttbar background.
To increase sensitivity, events are categorized by lepton decay channel ($\Pe\Pe$, $\Pgm\Pgm$, $\Pe\Pgm$), the number of AK4 jets ($\nj=4$, 5, 6, 7, ${\ge}8$), and how many of them are \PQb tagged ($\nb=2$, 3, ${\ge}4$).
The signal regions (SRs) contain seven or more jets, three or more of which are \PQb tagged by \textsc{DeepJet}, while categories containing fewer jets or exactly two \PQb-tagged jets serve as CRs.

The number of simulated \ttbar events with no jets from the hadronization of additional \PQb quarks is corrected to ensure consistency with the observed data.
The correction is determined from CRs which contain exactly one \PQb-tagged jet but satisfy all other criteria of the OSDL SRs and CRs included in the fit (\ie, two opposite-sign leptons, $\nj\ge4$, and $\HT>500\GeV$). The selected sample is depleted of \ttbar with extra \PQb jets.
The required correction is determined to be a scaling factor of $0.78 \pm 0.05$ (statistical uncertainty only), independent of lepton and jet kinematic properties.

\begin{figure}[htbp]
\centering
\includegraphics[width=0.8\columnwidth]{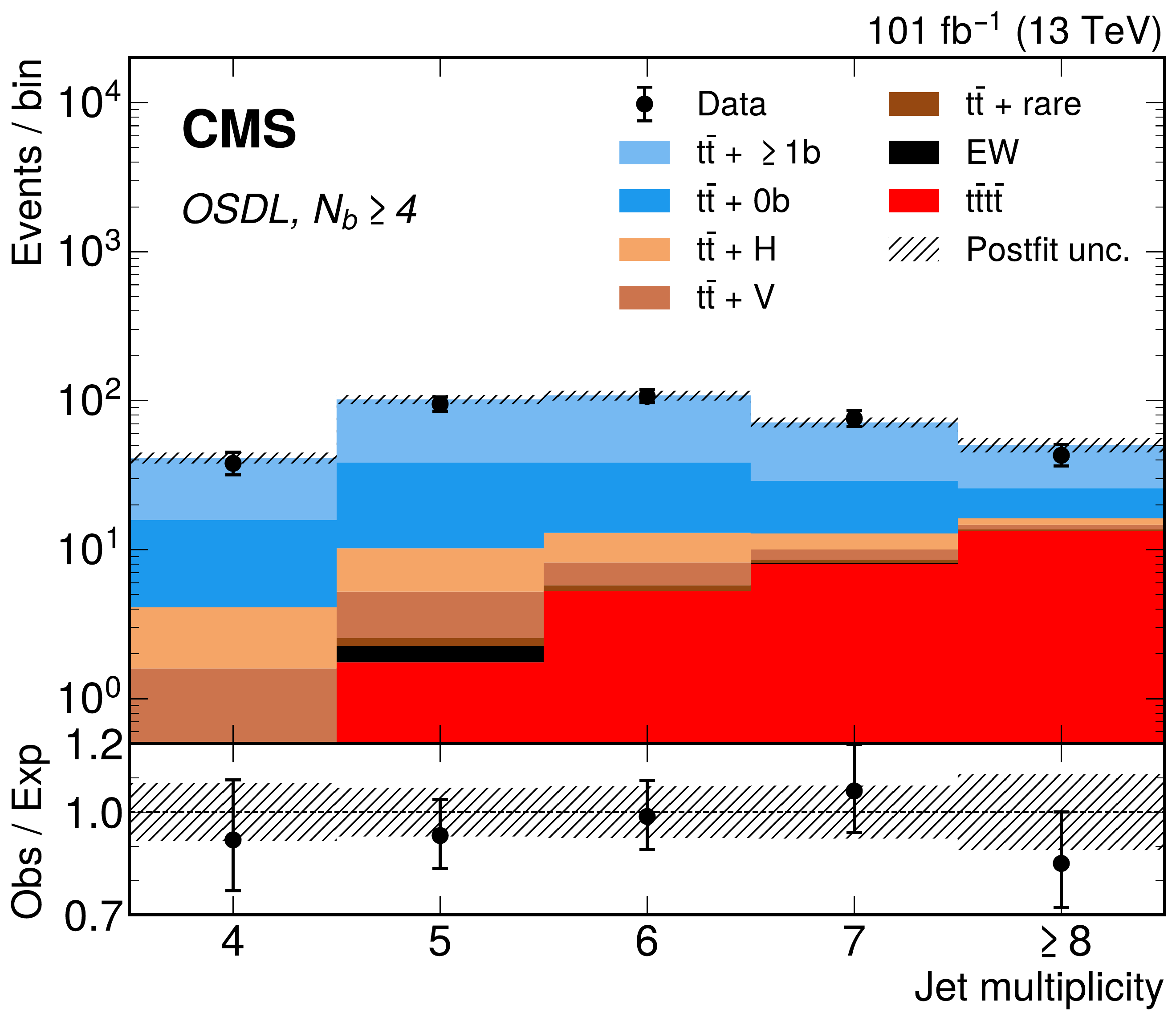}
\caption{\label{figure:OSDL}The jet multiplicity for $\nb\ge 4$ in the opposite-sign dilepton channel for the combined 2017--2018 dataset with dilepton decay categories combined.
Here, $\ttbar+{\geq} 1 \PQb$ refers to \ttbar events with at least one additional \PQb jet, $\ttbar+ 0 \PQb$ includes all other \ttbar events not produced in association with a boson, and EW refers to events that contain \PW and \PZ bosons but no top quarks.
The backgrounds and \tttt signal (derived from the fit described below) are shown as a stacked histogram.
The hatched bands correspond to the estimated total uncertainty after the fit.}
\end{figure}

The OSDL signal and background normalizations are obtained from a simultaneous binned maximum likelihood fit to all the categories, where a template is created for each jet and \PQb tag category, leptonic decay channel, and year.
Figure~\ref{figure:OSDL} shows the jet multiplicity distributions for the ${\ge}4$ \PQb tag categories after the fit to data.
The dominant background contribution changes with \PQb tag multiplicity.
With increasing \nb, backgrounds such as $\ttbar+{\geq} 1 \PQb$, \ttbar + \PH, and  \ttbar + V (V = \PW, \PZ) become more important. In the most sensitive categories, $\ttbar+{\geq} 1 \PQb$  becomes the dominant background.

\section{\label{sec:SingleLepton}Single-lepton final state}

Single-lepton events are required to have exactly one lepton with $\pt > 20\GeV$, at least four AK4 jets, at least two of which must be \PQb tagged using the \textsc{DeepCSV} algorithm, and $\HT>500\GeV$.
A BDT is trained using the \textsc{TMVA} package~\cite{Hocker:2007ht} to discriminate between signal and background in regions with large \nj and \nb.
A large number of possible BDT inputs are constructed, based on kinematic variables such as \pt, \PQb tagging discriminants, resolved \PQt tagging discriminants, mass, and angular separations of various objects and their combinations (\eg, jets, dijets, trijets, \PQb-tagged jets,  lepton--\PQb pairs, and $\bbbar$ pairs).
Information about the event topology is incorporated via event shape variables, such as centrality, planarity, sphericity, and the second Fox--Wolfram moment~\cite{FoxWolfram} calculated using all AK4 jets.
These potential inputs are evaluated based on whether they are well modeled in the simulation (using $\chi^2$, Kolmogorov-Smirnov, or Anderson-Darling goodness of fit tests) and whether they increase the BDT discrimination power. The final BDT classifier uses 40 inputs.
Signal and background samples are randomly divided into three equally populated parts: for training, for testing the performance, and for evaluating the classifier for the maximum likelihood fit.
The contributions from the dominant \ttbar background and all other SM processes are included in the training.
The training is optimized by targeting a region enriched in signal events by requiring $\nj\ge 6$ and $\nb\ge 2$.
The BDT is validated in the $\nb= 2$ region, which has a low signal-to-background ratio. Small corrections to the slope of the BDT distributions are derived in this region and applied to simulated events in the SR with $\nb\ge 3$.
The correction is verified to be valid in higher b-tag multiplicities, within uncertainties.

\begin{figure}[htbp]
\centering
\includegraphics[width=0.8\columnwidth]{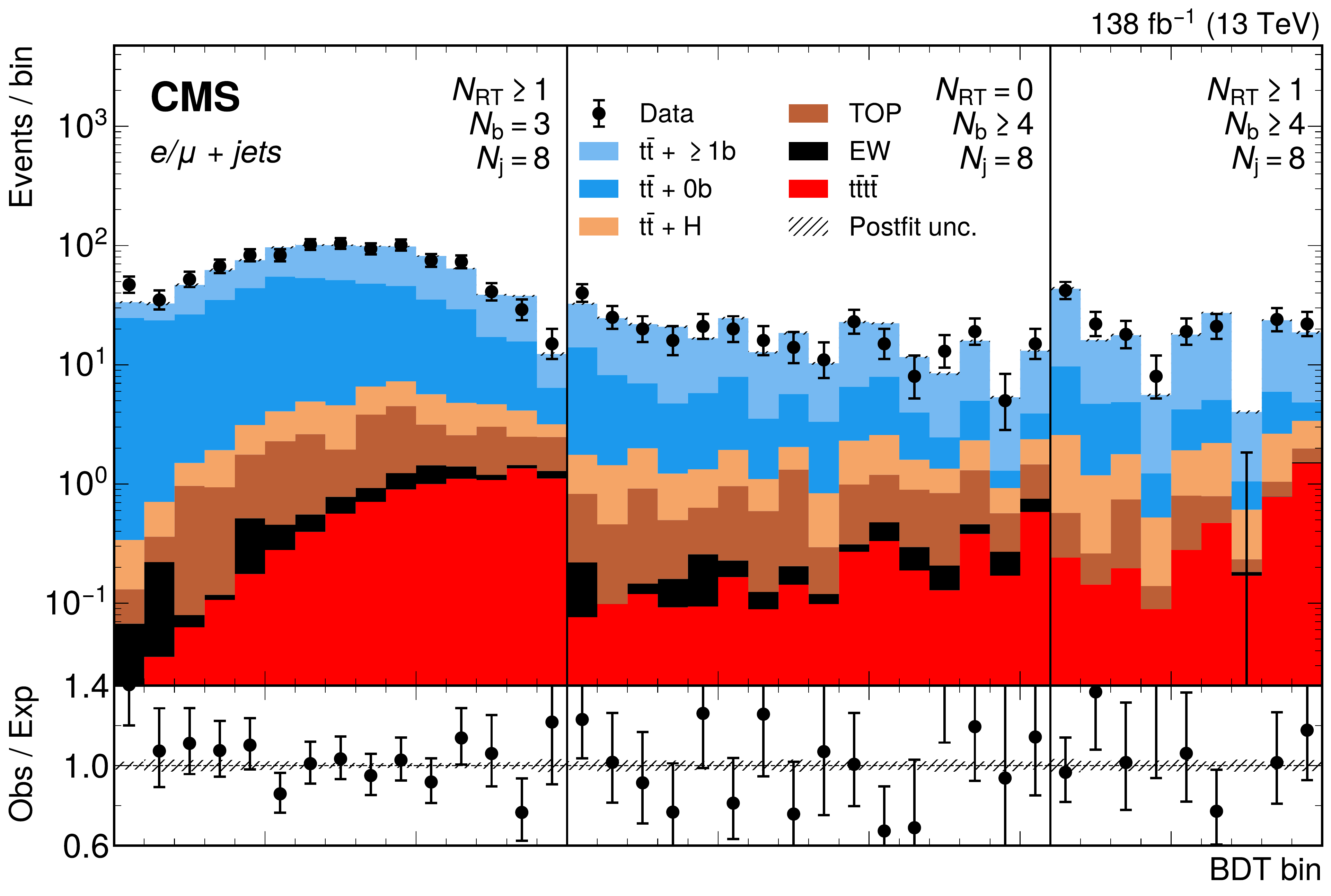}
\caption{\label{figure:SL}The distribution of the BDT discriminants for the 2016--2018 data set for three different categories in the combined single-electron and single-muon channels.
The three categories are defined by the number of resolved \PQt tags (\nrt), \PQb tags (\nb), and jets (\nj), selected as representative based on their sensitivity to signal.
Here, $\ttbar + {\geq} 1 \PQb$ refers to \ttbar events with at least one additional \PQb jet, while $\ttbar + 0 \PQb$ includes all other \ttbar events not produced in association with a boson.
The TOP grouping contains single top quark production along with the other \ttbar processes not explicitly shown, and EW refers to events that contain \PW and \PZ bosons but no top quarks.
The backgrounds and \tttt signal  (derived from the fit described below) are shown as a stacked histogram.
The hatched bands correspond to the estimated total uncertainty after the fit.
While the bins are shown to be equal width, they do not correspond to equal width in BDT value.}
\end{figure}

To maximize sensitivity to \tttt production, events are categorized based on lepton flavor (\Pe, \Pgm),
\nj (6, 7, 8, 9, ${\ge}10$), \nb (3, ${\ge}4$), and the number of resolved top quark candidates ($\nrt = 0$, ${\ge}1$).
The distribution of the BDT discriminant is used as the input to the final fit. The statistical interpretation is based on a simultaneous binned maximum likelihood fit to all categories.
The BDT distributions have variable bin sizes, chosen such that the statistical uncertainty in the total background in each bin is less than 30\%. 
Representative distributions from three different signal region categories are shown in Fig.~\ref{figure:SL}. 

\section{\label{sec:allhad}All-hadronic final state}

The all-hadronic SR is defined by requiring $\HT>700\GeV$, no prompt isolated leptons, at least nine AK4 jets of which at least three are \PQb tagged by \textsc{DeepJet}, and at least one resolved top quark candidate.
This selection results in a signal-to-background ratio of about $10^{-5}$ assuming the SM expectation for the signal.

Events in the SR are subdivided into 12 categories based on \nrt, the number of boosted top quarks (\nbt), and \HT as outlined in Table~\ref{tab:srdefinitions}.
The categorization by top quark tags defines three groups: $\nrt\ge2$, $\nrt=1$ and $\nbt\ge1$, and $\nrt=1$ and $\nbt=0$.
The first two groups are each further categorized into two ranges in \HT: 700--1100 and $>$1100\GeV. For the third group, there are six equally spaced bins in the range $700 < \HT <1300\GeV$, and two additional bins with \HT in the ranges 1300--1500 and $\ge$1500\GeV. 
The SR categories were chosen to optimize sensitivity to \tttt production. 

\begin{table}
\centering
\caption{\label{tab:srdefinitions} Definition of the 12 signal region (SR) categories for the all-hadronic channel, as a function of the number of resolved top quark candidates (\nrt), the number of boosted top quark candidates (\nbt), and \HT.} 
\begin{tabular}{ccc}
\nrt & \nbt & \HT range (\GeV)\\ \hline
1 & 0 & 700--800 \\
1 & 0 & 800--900 \\
1 & 0 & 900--1000 \\
1 & 0 & 1000--1100 \\
1 & 0 & 1100--1200 \\
1 & 0 & 1200--1300 \\
1 & 0 & 1300--1500 \\
1 & 0 & $\geq$1500 \\ \hline
1 & $\ge$1 & 700--1400 \\
1 & $\ge$1 & $\geq$1400 \\ \hline
$\ge$2 & $\ge$0 & 700--1100 \\
$\ge$2 & $\ge$0 & $\geq$1100 \\ \hline
\end{tabular}
\end{table}   

An event-level BDT is trained using \textsc{CatBoost}~\cite{prokhorenkova2019catboost} in each category of the SR to further distinguish between \tttt signal events and the dominant backgrounds originating from \ttbar and QCD multijet production, by exploiting differences in kinematic distributions of reconstructed objects.
The 20 optimized BDT input variables include \nj and \nb; the kinematic distributions of jets, \PQb-tagged jets, and \PQt-tagged candidates; and variables related to the angular distributions of jets. 

Techniques using CRs in data are employed to estimate the dominant backgrounds from QCD multijet and \ttbar + jet production, as described in the following text.
The ratio of the QCD multijet to the \ttbar + jet background is expected to be approximately 3:2 in the SR.

\begin{figure}[htbp]
\centering
\includegraphics[width=0.8\columnwidth]{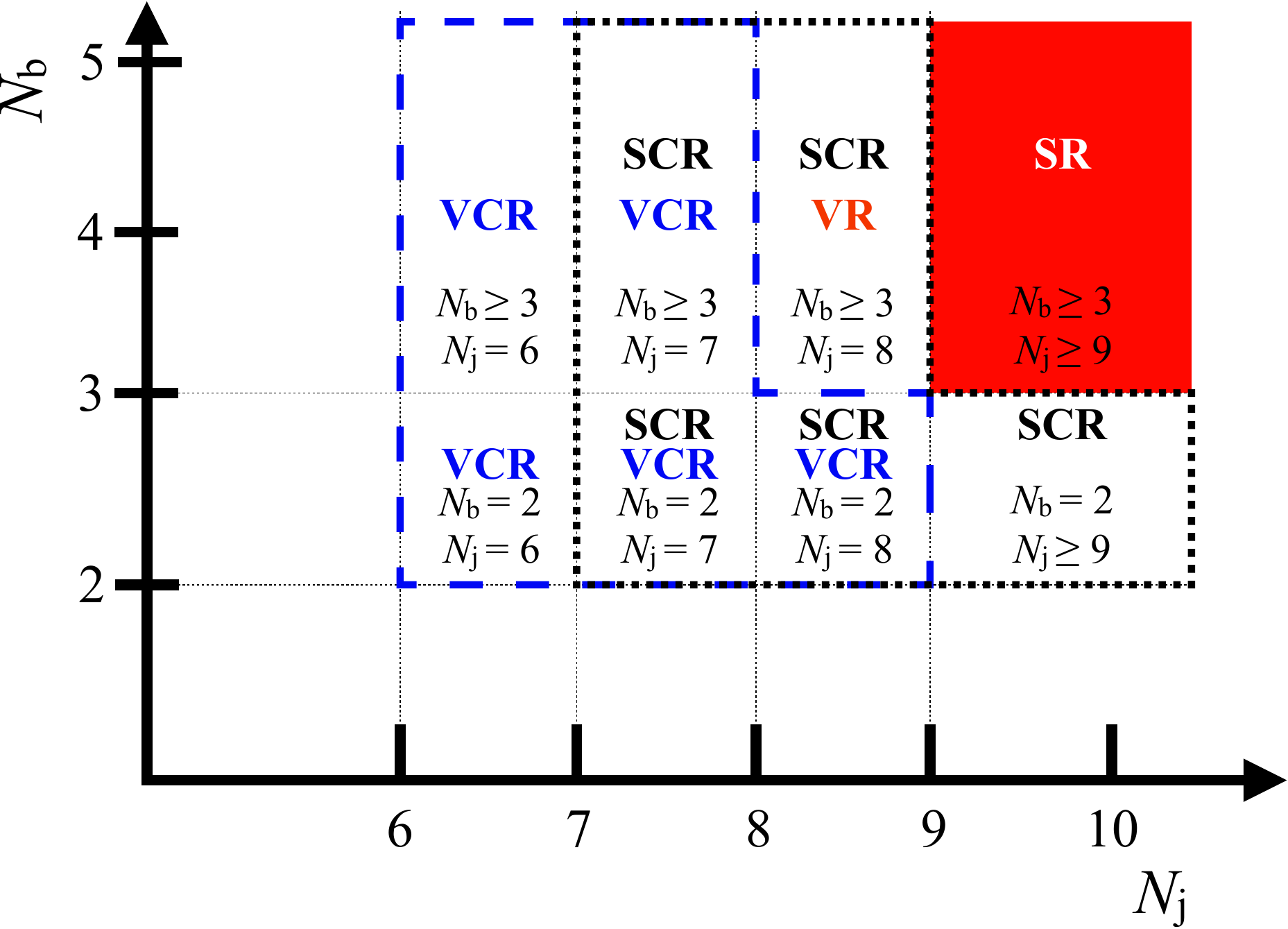}
\caption{\label{fig:abcdfig} Schematic showing the definitions of the signal region (SR), validation region (VR), signal control region (SCR), and validation control region (VCR) used in the all-hadronic analysis as a function of \nj and \nb.}
\end{figure}

Estimates of the absolute normalization of the background and the shape of the BDT distributions in the 12 SR categories are obtained from an extrapolation based on five CRs.
While the SR categories have  $\nj\ge 9$ and $\nb\ge 3$, these five CRs are defined to have
$(\nj,\nb) = (7,2)$, $(8,2)$, $({\ge}9,2)$, $(7,{\ge}3)$, and $(8,{\ge}3)$.
Figure~\ref{fig:abcdfig} illustrates how these control regions are related as a function of \nj and \nb. 
The absolute normalization of the background is estimated using an ''extended ABCD'' method~\cite{Choi:extendedABCD}, where the number of events in the SR is derived from the number of events in several independent CRs.
This method improves the accuracy of background yield estimates in cases where control variables are weakly correlated (such as \nj and \nb), compared to the traditional ABCD method in which only three CRs are used.
Specifically, this method is used to predict the number of \ttbar and QCD multijet events in the SR from the number of events observed in the five $(\nj,\nb)$ CRs, after subtracting the number of events from minor backgrounds. 

The shape of the BDT distribution in each SR is predicted using a DNN trained on the same five CRs that are used to estimate the absolute normalization.
A normalizing autoregressive flow~\cite{Choi:ABCDnn} is trained on the CRs to learn a bijective transformation of the \HT and BDT output distributions
between a source and a target.
The procedure is sensitive to statistical fluctuations in the source sample, so the input source used is a \ttbar sample with a large number of simulated events.
The target is the total \ttbar and multijet background (estimated by subtracting the other, simulated, background contributions from the data).
The QCD simulation is not included as an input source due to the very small number of simulated events in the regions of interest. 
In each of the CRs, the normalizing flow algorithm learns a transformation from the source distribution to the target distribution, before applying that transformation to the source distribution in the SR.
The validity of the method was demonstrated with various tests, including transformations between distributions from physics processes with large differences in shape.
The algorithm is trained regressively, starting from the less SR-like CRs (in terms of \nj and \nb) and including weights from previous training cycles in subsequent more SR-like training cycles.
For each data-taking period (2016, 2017, and 2018), the BDT output and \HT distributions are predicted simultaneously in the three SR \nrt and \nbt categories.
The predicted BDT distributions are then split according to the predicted \HT, resulting in predicted BDT output distributions binned in \HT for each of the 12 SR categories per data-taking period. These distributions are normalized to the absolute normalizations estimated from the ''extended ABCD'' extrapolation.

\begin{figure}[htbp]
  \centering
  \includegraphics[width=0.49\textwidth]{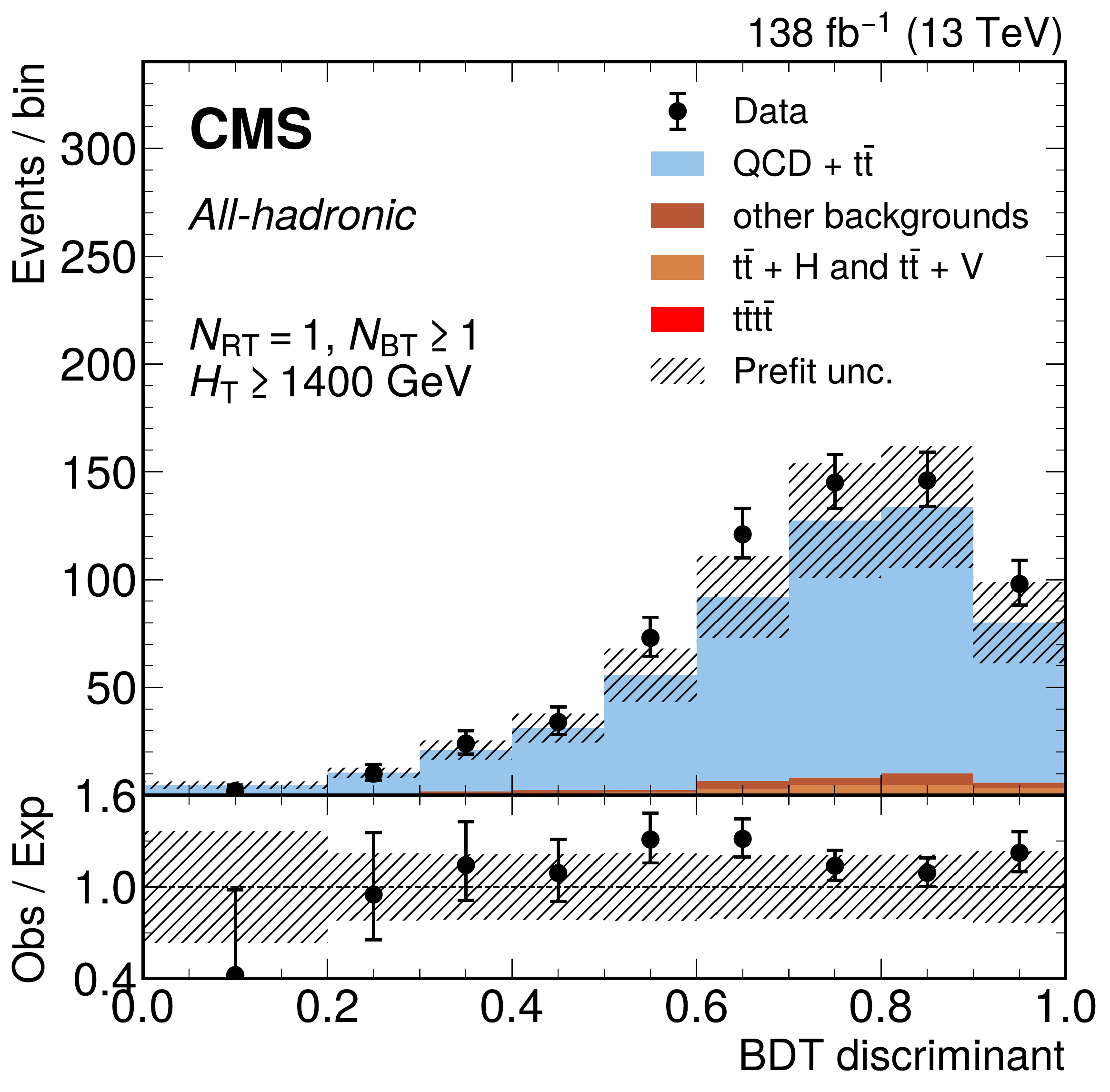}
  \includegraphics[width=0.49\textwidth]{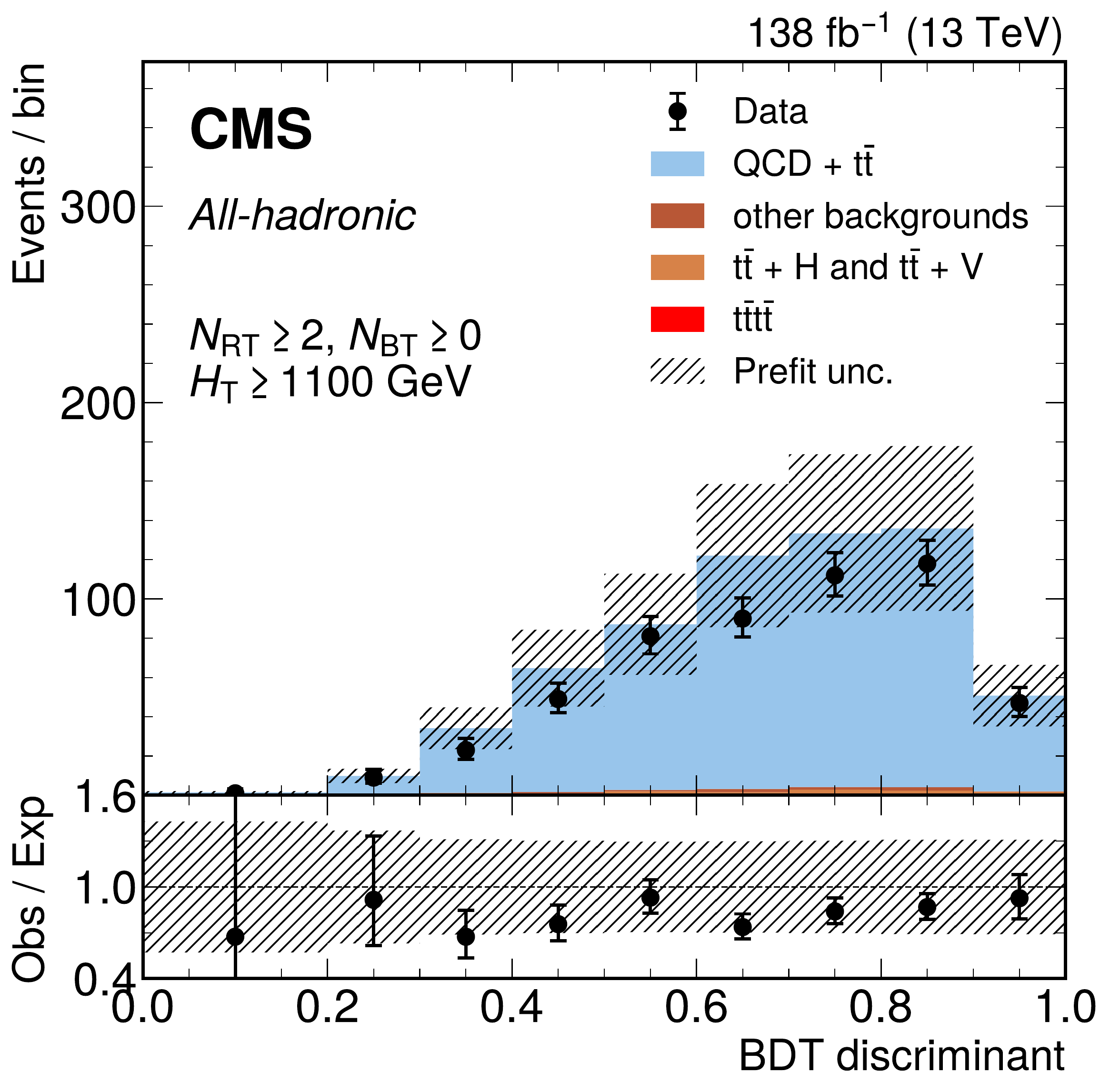}
  \caption{\label{fig:allhad_vr} The distribution of the BDT discriminants for the full 2016--2018 data set in the all-hadronic channel for two VRs. The two sample VRs are defined by $\nrt=1$, $\nbt\ge1$, $\HT>1400\GeV$ (\cmsLeft), and $\nbt\ge2$, $\HT>1100\GeV$ (\cmsRight). The background from QCD multijet and \ttbar production is derived from control regions in the data. Estimates for the \tttt signal and other backgrounds are shown using simulated samples. The hatched bands correspond to the estimated total uncertainty.}
\end{figure}

The predictions are checked in validation regions (VRs) defined in parallel to those for the SR, but with $\nj=8$, $\nb\ge3$ and extrapolating from CRs in which $(\nj,\nb)= (6,2)$, $(7,2)$, $(8,2)$, $(6,{\ge}3)$, and $(7,{\ge}3)$. The VRs and their associated CRs are split into the same \nrt, \nbt, and \HT categories as the SRs and their associated CRs as detailed in Table~\ref{tab:srdefinitions}.
Figure~\ref{fig:allhad_vr} shows BDT disciminant distributions in two VRs, which demonstrate the robustness of this method.

For each category, the difference between the total number of predicted and observed events in the corresponding VR is taken as a normalization uncertainty.
This is added in quadrature to a second normalization uncertainty estimated from the bin-to-bin variations within the VR.
Shape uncertainties are also estimated for each category, characterized as a linear scaling of the BDT discriminant for each event.
The maximum scaling required for the predicted and observed distribution shapes to agree within the uncertainty in all bins in any VR is $\pm 3$\%.
These uncertainties are considered to be correlated for any \HT categories that use the same DNN training.

\begin{figure}[htbp]
\centering
\includegraphics[width=0.49\textwidth]{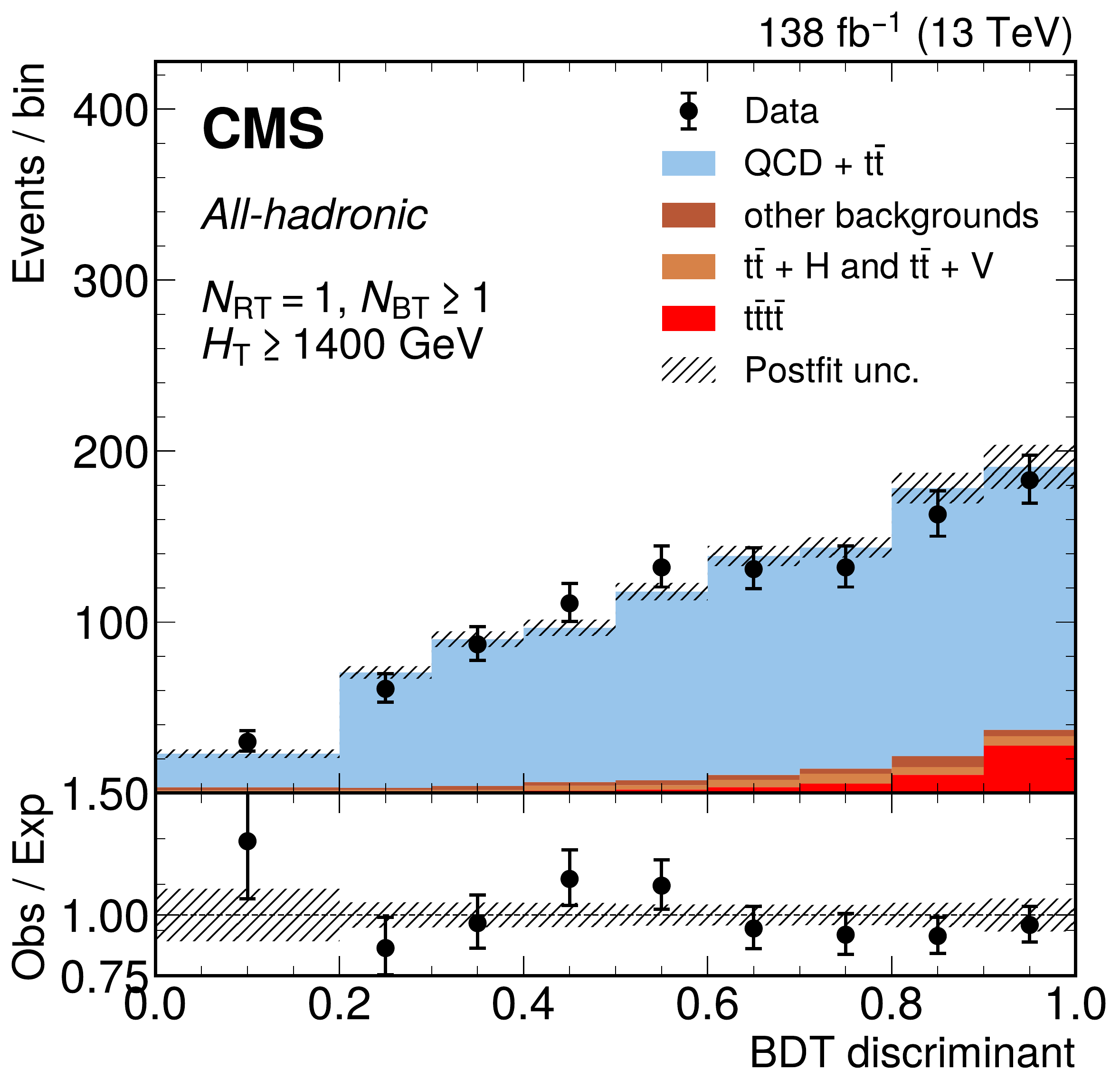}
\includegraphics[width=0.49\textwidth]{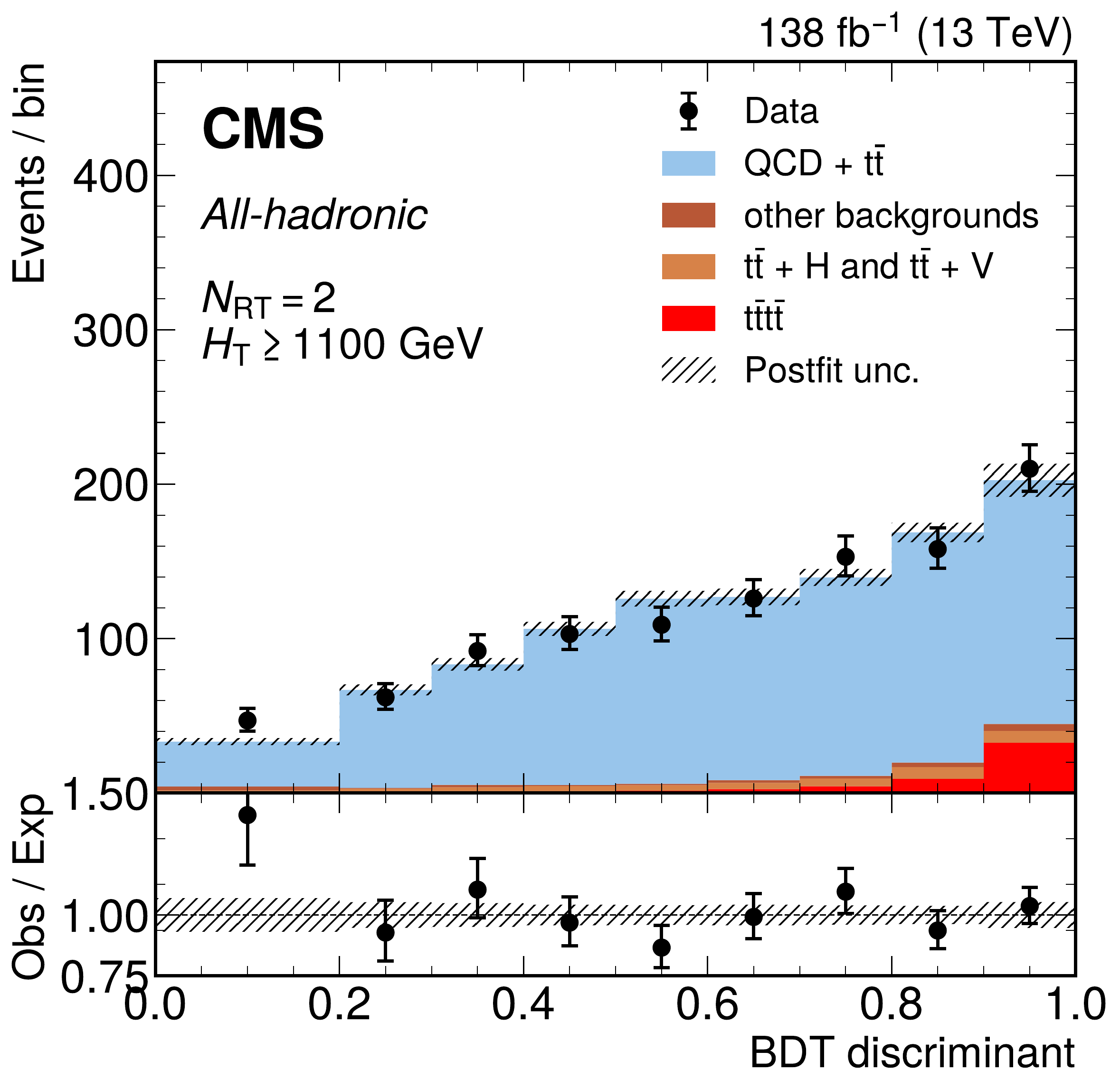}
\caption{\label{fig:allhad} The distribution of the BDT discriminants for the full 2016--2018 data set in the all-hadronic channel. The two most sensitive SR categories are shown, defined by $\nrt=1$, $\nbt\ge1$, $\HT>1400\GeV$ (\cmsLeft), and $\nbt\ge2$, $\HT>1100\GeV$ (\cmsRight). The background from QCD multijet and \ttbar production is derived from control regions in the data. Estimates for the \tttt signal and other backgrounds are shown using simulated samples. The hatched bands correspond to the estimated total uncertainty after the fit.}
\end{figure}

Other backgrounds, originating from the associated production of \ttbar with {\PW}, {\PZ}, or \PH bosons, are estimated from simulation.
Figure~\ref{fig:allhad} shows the BDT output distributions for the two SR categories that are most sensitive to the \tttt signal.
The distribution of the BDT discriminant in each category is used as the input to the final fit.

\section{\label{systematics}Statistical procedure and systematic uncertainties}

An estimate of the \tttt production significance and the measurement of the cross section is obtained by a likelihood fit, based on the procedure described in Refs.~\cite{Read:2002hq,Junk:1999kv}. The fit incorporates the experimental and theoretical uncertainties as nuisance parameters, assuming that shared systematic uncertainties between analyses are correlated.
Different \PQb tagging algorithms are treated as uncorrelated; studies confirm that this choice has no significant effect on the fit results.
The significance of signal events is obtained from a comparison to the background-only hypothesis assuming the asymptotic regime as detailed in Ref.~\cite{Cowan:2010js}.

A likelihood fit is performed to the all-hadronic, single-lepton, and OSDL results first, and then a second likelihood fit is performed including previously published results.
Uncertainties in both the normalizations and the shapes of the discriminant distributions are considered for all background processes.
For the signal, the uncertainties are applied to the predicted shapes and normalizations of the discriminant distributions while the cross section is unconstrained. 
Systematic uncertainties that come from the same experimental or theoretical source are treated consistently between results.

The sensitivity of the analysis is limited primarily by the small size of the sample, which contributes an uncertainty of 22\% in the measured signal strength.
Approximately 400 nuisance parameters representing systematic uncertainties are considered; the two with the largest individual effects are the theoretical uncertainty in the {\PQt}{\PQt}\PH cross section (4.6\% impact on signal strength) and the uncertainty in the modeling of \ttbar with additional \PQb~jets (3.7\%).
The background estimation in the all-hadronic final state contributes up to 2.7\% per SR category, dominated by statistical fluctuations in the CRs.
The jet energy scale contributes up to 2.4\% (depending on the year and channel), renormalization and factorization scales 2.1\%, and leptonic fake rates 1.9\%.
The largest components of the \PQb~tagging and light quark mistagging efficiency uncertainties each contribute 1.8\%.

Further nuisance parameters with smaller individual contributions include those relating to lepton reconstruction and trigger efficiencies, the input cross section for \ttbar + additional \PQb~quark production, jet energy resolution, matrix element to parton shower matching, the modeling of the resolved \PQt tagging in the single-lepton channel, PDFs, pileup, the theoretical uncertainty in the \ttbar cross section, the theoretical modeling of initial-state radiation in \ttbar production, and the delivered luminosity.
The total systematic uncertainty, considering the effects of all nuisance parameters, is 17\%.

\section{Results \label{sec:results}}

\begin{figure}[htbp]
        \centering
        \includegraphics[width=0.8\columnwidth]{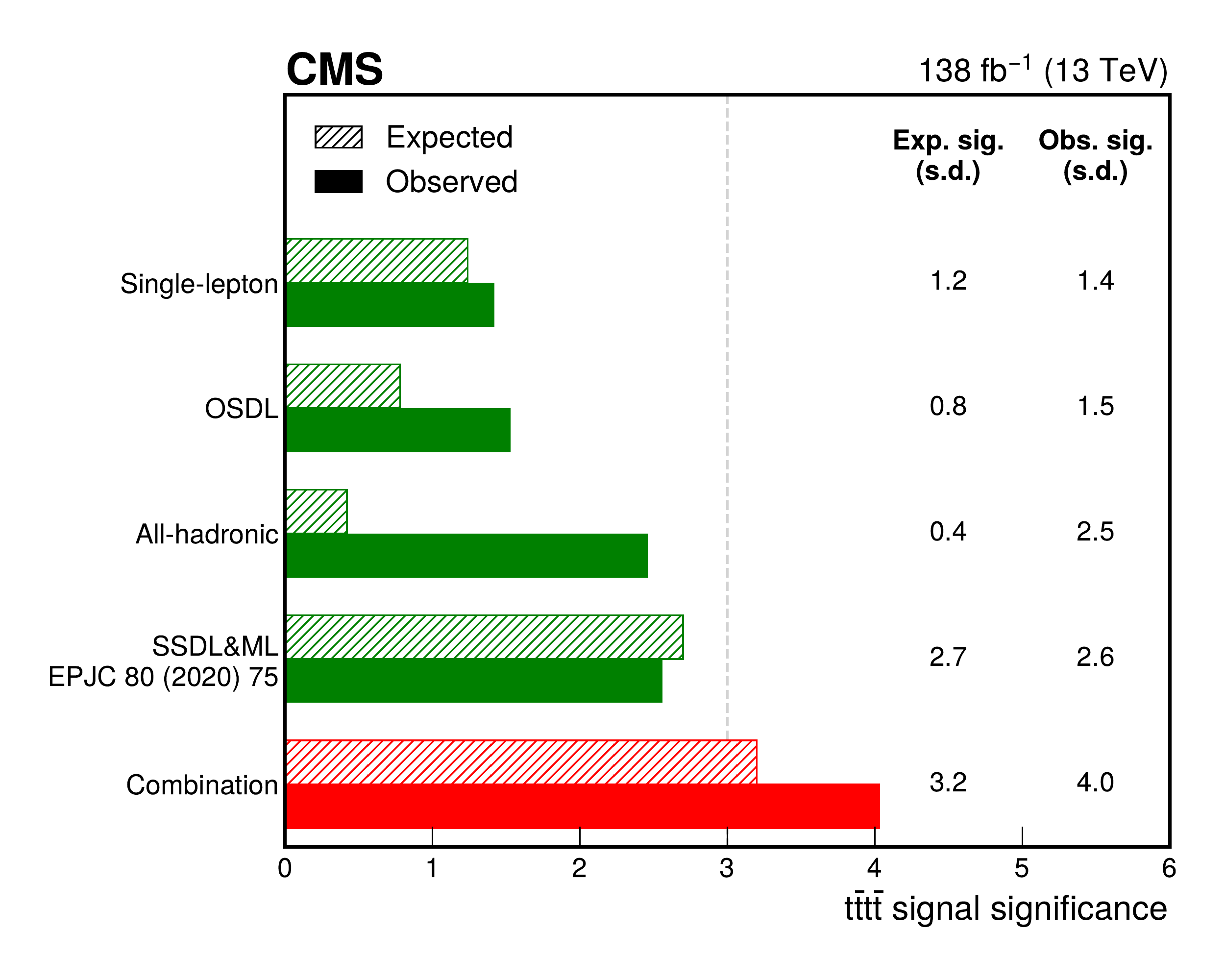}
        \caption{Expected and observed significance (in standard deviations) for \tttt production from each final state and the combination with previous CMS results~\cite{Sirunyan:2019wxt,Sirunyan:2019nxl}. The same-sign dilepton and multilepton (SSDL\&ML) final state results are from Ref.~\cite{Sirunyan:2019wxt}.}
        \label{fig:significance_summary}
\end{figure}

Table~\ref{tab:combination_summary} shows the fitted values of the signal strength (the ratio of the measured cross section to the prediction), the measured cross section, and the expected and observed significance of \tttt production from 2017--2018 data in the OSDL channel, and 2016--2018 data in the single-lepton and all-hadronic channels.
The signal strength is calculated with all systematic uncertainties, including all theoretical uncertainties.
The cross section measurement is performed with all systematic uncertainties except theoretical uncertainties that affect the rate of the signal process. 
The table also includes the combination of these new results with the CMS OSDL analysis of 2016 data~\cite{Sirunyan:2019nxl},
and the same-sign dilepton and multilepton analysis of the 2016--2018 data~\cite{Sirunyan:2019wxt}, following the procedure
described in Refs.~\cite{Read:2002hq,Junk:1999kv}.
The expected and observed significances for each final state and the combination are also shown in Fig.~\ref{fig:significance_summary}.
The combination of results described in this Letter gives a signal strength of $2.5 \pm 0.5 \stat \pm 0.5 \syst$
and a measured cross section of $36\pm 7 \stat \phantom{I}^{+10}_{-8}\syst$,
providing evidence for
\tttt production with an observed significance of 3.9 standard deviations (1.5 expected).
The full combination of all channels for the CMS 2016--2018 data set gives a signal strength of $1.4 \pm 0.3 \stat \pm 0.2\syst$ and a measured
cross section of $17\pm 4\stat\pm 3\syst$, increasing the observed significance to 4.0 standard deviations (3.2 expected).
This is the most sensitive \tttt analysis to date.

\begin{table*}[htbp]
\centering
\topcaption{Measured signal strength ($\mu = \sigma_{\tttt}/\sigma_{\tttt}^{\mathrm{SM}}$ where $\sigma_{\tttt}^{\mathrm{SM}} = 12\unit{fb}$), corresponding cross section (in fb), and the expected and observed significance (in standard deviations) for \tttt production from all analysis channels. This table shows production from each analysis channel in this Letter, the combination of those channels, the results from previously published results, and the full combination of all CMS 2016--2018 results.}
\renewcommand{\arraystretch}{1.3}
\begin{tabular}{ccccccccc}
Analysis & \multicolumn{3}{c}{Signal strength ($\mu$)} & \multicolumn{3}{c}{Cross section (fb)} & \multicolumn{2}{c}{Significance (s.d.)} \\
&  & (stat.) & (syst.) & & (stat.) & (syst.) &  Exp.  &  Obs. \\ \hline
OSDL (2017+2018)       &  2.8 & $\pm 1.0$          & $\!^{+1.9}_{-1.2}$ & 33 & $\pm 12$         & $\!^{+15}_{-14}$ & 0.6   & 1.8   \\ 
Single-lepton          &  1.2 & $\!^{+0.7}_{-0.6}$ & $\pm 0.6$          & 15 & $\pm 8$          & $\!^{+10}_{-7 }$ & 1.2   & 1.4   \\ 
All-hadronic           &  5.8 & $\pm 1.4$          & $\pm 2.0$          & 70 & $\pm 17$         & $\!^{+25}_{-23}$ & 0.4   & 2.5   \\ 
\textbf{Combination of above}   &  2.5 & $\pm 0.5$          & $\pm 0.5$          & 36 & $\pm 7$          & $\!^{+10}_{-8 }$ & 1.5   & 3.9   \\ [\cmsTabSkip]
SSDL\&ML (2016--2018)~\cite{Sirunyan:2019wxt}   &  1.0 & $\pm 0.4$          & $\!^{+0.3}_{-0.2}$ & 13 & $\!^{+5 }_{-4 }$ & $\pm 3$          & 2.7   & 2.6   \\ 
OSDL (2016)~\cite{Sirunyan:2019nxl}            & $-0.2$ & $\!^{+1.7}_{-1.5}$ & $\pm 1.5$          & $-2$ & $\!^{+20}_{-18}$ & $\pm 18$         & 0.4   & 0     \\ [\cmsTabSkip] 
\textbf{Full combination} &  1.4 & $\pm 0.3$ & $\pm 0.2$ & 17 & $\pm 4$ & $\pm 3$ & 3.2 & 4.0 \\ 
\end{tabular}
\label{tab:combination_summary}
\end{table*}

\section{Summary}
We have measured the cross section for the simultaneous production of four top quarks (\tttt) in proton-proton collisions. The data were collected by the CMS experiment at the LHC in 2016--2018, and correspond to an integrated luminosity of up to 138\fbinv at a center-of-mass energy of 13\TeV. The all-hadronic final state has been studied for the first time in a \tttt production analysis, using a background estimation strategy based on a deep neural network trained using control regions in data. Final states with one lepton (electron or muon), or two opposite-sign leptons have also been analyzed. The observed and expected significances obtained from the combination of the new analyses described here are 3.9 and 1.5 standard deviations, respectively. When combined with published CMS results in other final states, the significances increase to 4.0 (observed) and 3.2 (expected) standard deviations. This is a significant improvement compared to previous CMS results and the first CMS evidence for \tttt production with a significance above three standard deviations.

\begin{acknowledgments}
We congratulate our colleagues in the CERN accelerator departments for the excellent performance of the LHC and thank the technical and administrative staffs at CERN and at other CMS institutes for their contributions to the success of the CMS effort. In addition, we gratefully acknowledge the computing centers and personnel of the Worldwide LHC Computing Grid and other centers for delivering so effectively the computing infrastructure essential to our analyses. Finally, we acknowledge the enduring support for the construction and operation of the LHC, the CMS detector, and the supporting computing infrastructure provided by the following funding agencies: BMBWF and FWF (Austria); FNRS and FWO (Belgium); CNPq, CAPES, FAPERJ, FAPERGS, and FAPESP (Brazil); MES and BNSF (Bulgaria); CERN; CAS, MoST, and NSFC (China); MINCIENCIAS (Colombia); MSES and CSF (Croatia); RIF (Cyprus); SENESCYT (Ecuador); MoER, ERC PUT and ERDF (Estonia); Academy of Finland, MEC, and HIP (Finland); CEA and CNRS/IN2P3 (France); BMBF, DFG, and HGF (Germany); GSRI (Greece); NKFIH (Hungary); DAE and DST (India); IPM (Iran); SFI (Ireland); INFN (Italy); MSIP and NRF (Republic of Korea); MES (Latvia); LAS (Lithuania); MOE and UM (Malaysia); BUAP, CINVESTAV, CONACYT, LNS, SEP, and UASLP-FAI (Mexico); MOS (Montenegro); MBIE (New Zealand); PAEC (Pakistan); MES and NSC (Poland); FCT (Portugal);  MESTD (Serbia); MCIN/AEI and PCTI (Spain); MOSTR (Sri Lanka); Swiss Funding Agencies (Switzerland); MST (Taipei); MHESI and NSTDA (Thailand); TUBITAK and TENMAK (Turkey); NASU (Ukraine); STFC (United Kingdom); DOE and NSF (USA).
  
\hyphenation{Rachada-pisek} Individuals have received support from the Marie-Curie program and the European Research Council and Horizon 2020 Grant, contract Nos.\ 675440, 724704, 752730, 758316, 765710, 824093, 884104, and COST Action CA16108 (European Union); the Leventis Foundation; the Alfred P.\ Sloan Foundation; the Alexander von Humboldt Foundation; the Belgian Federal Science Policy Office; the Fonds pour la Formation \`a la Recherche dans l'Industrie et dans l'Agriculture (FRIA-Belgium); the Agentschap voor Innovatie door Wetenschap en Technologie (IWT-Belgium); the F.R.S.-FNRS and FWO (Belgium) under the ``Excellence of Science -- EOS" -- be.h project n.\ 30820817; the Beijing Municipal Science \& Technology Commission, No. Z191100007219010; the Ministry of Education, Youth and Sports (MEYS) of the Czech Republic; the Hellenic Foundation for Research and Innovation (HFRI), Project Number 2288 (Greece); the Deutsche Forschungsgemeinschaft (DFG), under Germany's Excellence Strategy -- EXC 2121 ``Quantum Universe" -- 390833306, and under project number 400140256 - GRK2497; the Hungarian Academy of Sciences, the New National Excellence Program - \'UNKP, the NKFIH research grants K 124845, K 124850, K 128713, K 128786, K 129058, K 131991, K 133046, K 138136, K 143460, K 143477, 2020-2.2.1-ED-2021-00181, and TKP2021-NKTA-64 (Hungary); the Council of Science and Industrial Research, India; the Latvian Council of Science; the Ministry of Education and Science, project no. 2022/WK/14, and the National Science Center, contracts Opus 2021/41/B/ST2/01369 and 2021/43/B/ST2/01552 (Poland); the Funda\c{c}\~ao para a Ci\^encia e a Tecnologia, grant CEECIND/01334/2018 (Portugal); MCIN/AEI/10.13039/501100011033, ERDF ``a way of making Europe", and the Programa Estatal de Fomento de la Investigaci{\'o}n Cient{\'i}fica y T{\'e}cnica de Excelencia Mar\'{\i}a de Maeztu, grant MDM-2017-0765 and Programa Severo Ochoa del Principado de Asturias (Spain); the Chulalongkorn Academic into Its 2nd Century Project Advancement Project, and the National Science, Research and Innovation Fund via the Program Management Unit for Human Resources \& Institutional Development, Research and Innovation, grant B05F650021 (Thailand); the Kavli Foundation; the Nvidia Corporation; the SuperMicro Corporation; the Welch Foundation, contract C-1845; and the Weston Havens Foundation (USA).

\end{acknowledgments}

\bibliography{auto_generated}
\cleardoublepage \appendix\section{The CMS Collaboration \label{app:collab}}\begin{sloppypar}\hyphenpenalty=5000\widowpenalty=500\clubpenalty=5000
\cmsinstitute{Yerevan Physics Institute, Yerevan, Armenia}
{\tolerance=6000
A.~Tumasyan\cmsAuthorMark{1}\cmsorcid{0009-0000-0684-6742}
\par}
\cmsinstitute{Institut f\"{u}r Hochenergiephysik, Vienna, Austria}
{\tolerance=6000
W.~Adam\cmsorcid{0000-0001-9099-4341}, J.W.~Andrejkovic, T.~Bergauer\cmsorcid{0000-0002-5786-0293}, S.~Chatterjee\cmsorcid{0000-0003-2660-0349}, K.~Damanakis\cmsorcid{0000-0001-5389-2872}, M.~Dragicevic\cmsorcid{0000-0003-1967-6783}, A.~Escalante~Del~Valle\cmsorcid{0000-0002-9702-6359}, P.S.~Hussain\cmsorcid{0000-0002-4825-5278}, M.~Jeitler\cmsAuthorMark{2}\cmsorcid{0000-0002-5141-9560}, N.~Krammer\cmsorcid{0000-0002-0548-0985}, L.~Lechner\cmsorcid{0000-0002-3065-1141}, D.~Liko\cmsorcid{0000-0002-3380-473X}, I.~Mikulec\cmsorcid{0000-0003-0385-2746}, P.~Paulitsch, J.~Schieck\cmsAuthorMark{2}\cmsorcid{0000-0002-1058-8093}, R.~Sch\"{o}fbeck\cmsorcid{0000-0002-2332-8784}, D.~Schwarz\cmsorcid{0000-0002-3821-7331}, M.~Sonawane\cmsorcid{0000-0003-0510-7010}, S.~Templ\cmsorcid{0000-0003-3137-5692}, W.~Waltenberger\cmsorcid{0000-0002-6215-7228}, C.-E.~Wulz\cmsAuthorMark{2}\cmsorcid{0000-0001-9226-5812}
% ...
\par}
\vskip\cmsinstskip
\dag:~Deceased\\
$^{1}$Also at Yerevan State University, Yerevan, Armenia\\
$^{2}$Also at TU Wien, Vienna, Austria\\
% ...
\end{sloppypar}
\end{document}